\documentclass[10pt,
               aps,
               pre,
               twocolumn,
               showpacs, 
               superscriptaddress]{revtex4-2}
\usepackage[utf8]{inputenc}
\usepackage{graphicx}

\usepackage[dvipsnames,table]{xcolor}
\usepackage[caption=false]{subfig}
\usepackage{amsmath, amsfonts, amssymb}
\usepackage{physics}
\usepackage[linktocpage=true,
  colorlinks=true, 
  pdfborder={0 0 0},
  linkcolor=blue,
  citecolor=blue,
  filecolor=yellow,
  urlcolor=blue,
  bookmarks,
  pdfauthor={},
]{hyperref}
\usepackage{tikz}
\graphicspath{{images/}}

\newcommand{\msd}{\ensuremath{\langle \Delta r^2(t) \rangle}}
\newcommand{\vdiv}{\ensuremath{\vb*{\nabla_v} \vdot}}
\newcommand{\vgrad}{\grad_{\vb*{v}}}
\newcommand{\vlaplacian}{\ensuremath{\laplacian_v}}
\newcommand{\profile}{\vb*{p}_\phi(\vb*{r})}
\newcommand{\rll}{\ensuremath{r_\parallel}}
\newcommand{\rT}{\ensuremath{r_\perp}}
\newcommand{\vll}{\ensuremath{v_\parallel}}
\newcommand{\vT}{\ensuremath{v_\perp}}

\makeatletter
\newcommand*{\rom}[1]{\expandafter\romannumeral #1}
\makeatother

\begin{document}

\title{Inertial chiral active Brownian particle: Transition from Gaussian to platykurtic distribution}
\author{M Muhsin}
\thanks{These authors contributed equally to this work.}
\affiliation{Department of Physics, University of Kerala, Kariavattom, Thiruvananthapuram-$695581$, India}
\affiliation{Institut f\"ur Physik, Otto-von-Guericke-Universit\"at Magdeburg, Universit\"atsplatz 2, 39106 Magdeburg, Germany}

\author{S Deion}
\thanks{These authors contributed equally to this work.}
\affiliation{Department of Physics, University of Kerala, Kariavattom, Thiruvananthapuram-$695581$, India}

\author{M Sahoo}
\email{jolly.iopb@gmail.com}
\affiliation{Department of Physics, University of Kerala, Kariavattom, Thiruvananthapuram-$695581$, India}

\date{\today}

\begin{abstract}
We investigate the dynamics of an inertial chiral active Brownian particle in the presence of a harmonic confinement. Through numerical simulation, we observe that when the harmonic frequency becomes comparable to the chiral frequency, the position distribution transitions from a Gaussian to a platykurtic distribution, corresponding to short tails with a nearly uniform probability near the minimum of the potential. This result is further confirmed by analyzing the kurtosis of the position of the particle as a function of harmonic frequency, which exhibits a dip when the harmonic frequency matches the chiral frequency.
At the same time, the steady state mean square displacement (MSD) shows a non-monotonic feature with the harmonic frequency and shows a maximum only when the harmonic frequency is of the same order as the chiral frequency. 
In the rotational overdamped limit of the same model, we have calculated the exact expression for kurtosis, steady state MSD and find that the qualitative behavior remains the same. Kurtosis still exhibits a dip in the matching of chiral and harmonic frequencies, but the feature is less pronounced with a higher minimum.
These findings might be relevant for controlling the transport and spatial distribution of chiral microswimmers in optical or acoustic traps, where confinement can be tuned to optimize particle distribution.
\end{abstract}

\maketitle

\section{INTRODUCTION}
Active matter encompasses systems composed of self-propelled entities that continuously convert energy from their surroundings into directed motion, thereby maintaining them inherently out of equilibrium~\cite{Ramaswamy2017active, volpe2022active}. This class of materials spans a vast range, from microscopic biological systems such as motile bacteria~\cite{son2013bacteria,aranson2022bacterial,copeland2009bacterial}, cytoskeletal filaments~\cite{banerjee2011_cytoskeletal}, and cellular assemblies~\cite{MacKintosh2010_cellular} to engineered microswimmers~\cite{howse2007self}, synthetic Janus particles~\cite{walther2013janus}, and colloids~\cite{zottl2016}. The Active Brownian Particle (ABP) model~\cite{hagen2011brownian, cates2013when, stenhammar2014phase, romanczuk2012active, lowen2020inertial, kanaya2020steady, solon2015pressure, caprini2021collective, caprini2020hidden, buttinoni2013dynamical, bialke2015negative, mandal2019motility} serves as one of the most fundamental frameworks to describe the dynamics of such systems, combining self-propulsion with rotational diffusion in a minimal setting. Despite its simplicity, the model captures many key nonequilibrium features observed in experiments, including enhanced diffusion~\cite{howse2007self,hagen2011brownian}, spontaneous accumulation near boundaries~\cite{Wensink2008_boundary_accumulation}, and motility-induced phase separation~\cite{Tailleur2008_phase_sepration,fily2012athermal}. During the past decade, ABPs have become a cornerstone in the theoretical study of active matter, offering a bridge between the microscopic mechanisms of self-propulsion and the emergent collective behaviors that define these fascinating systems~\cite{marchetti2013hydrodynamics,bechinger2016active,cates2015motility}.

A natural extension of the ABP framework arises when the self-propulsion direction acquires intrinsic rotation, giving rise to chiral active particles (CAPs)~\cite{liebchen2022chiral, chan2024chiral, sevilla201diffusion, caprini2019activechiral, Caprini2023chiral, pagonabarraga2019activity, bickmann2022analytical, ai2016ratchet, li2022chiral, kummel2013circular}. Chirality introduces a persistent circular motion which profoundly modifies the transport and accumulation properties, especially under confinement. 
Studies of CAPs in restricted geometries have uncovered several distinctive features~\cite{Berke2008hydrodynamic,Volpe2014simulation,liebchen2017collective}. For example, near confining boundaries, chiral activity can generate persistent surface currents and enhanced bulk accumulation~\cite{caprini2019activechiral}. In non-radial trapping potentials, chirality can break the parity symmetry, can generate direction reversal of rotational motion~\cite{deion2025chiralactive}, and introduce correlations between orthogonal spatial components~\cite{Caprini2023chiral}.
Furthermore, under confinement, chirality can produce re-entrant transitions of kurtosis governed by dimensionality~\cite{pattanayak2025chirality}.
Beyond single-particle behavior, collective chiral motion can induce several phenomena such as pattern formation~\cite{liebchen2017collective, levis2018micro-flock}, self-reverting vortices~\cite{caprini2024self-reverting}, glassy transitions~\cite{ai2024rotational}, and so on. Nevertheless, most theoretical and numerical investigations have focused on the overdamped regime, leaving important aspects of underdamped CAP behavior underexplored. Incorporating inertia becomes essential when dealing with larger active particles, vibrobots, or granular active matter, where the impact of inertia cannot be neglected~\cite{scholz2018inertial, dauchot2019dynamics, debasis2018boundaries, gutierrez2025time-dependent, herrera2021maxwell}. 
In the recent progress of active matter, the interplay between inertia and chirality reveals qualitatively new dynamical features. In particular, inertial extensions of a chiral active Brownian particle exhibit a re-entrant transition with mass, and the activity is suppressed by increasing chirality~\cite{Pattanayak2025InertiaChirality}. Similarly, an inertial chiral particle exhibits chirality-induced transient self-trapping~\cite{sahala2025self}.
The interplay between intrinsic chirality, confinement, and inertial effects is therefore expected to generate qualitatively new steady states and transport characteristics beyond the standard overdamped description.

In this work, we consider an inertial chiral ABP confined in a harmonic potential and investigate the transport properties of both its transient and steady-state behavior. When inertia is present in both translational and rotational dynamics, we observe a resonant-like effect, where the steady state mean square displacement (MSD) of the particle is enhanced when the chiral and harmonic frequencies are of the same order. We examine this phenomenon in detail by analyzing the position probability distribution, kurtosis, and orientation profile. When both the harmonic and chiral frequencies are approximately equal in magnitude, the position distribution deviates from the Gaussian form and evolves into a platykurtic distribution with comparatively shorter tails. Continually, the kurtosis displays a pronounced dip at the point where both frequencies match. In the same regime, the orientation profile develops a circular structure, reflecting the strong coupling between chirality, activity, and confinement. We further focus our attention on the case where the orientational dynamics is overdamped while retaining inertia in the translational motion. In this limit, we exactly derive analytical expressions for the steady state MSD and kurtosis of the particle, and compare it with the simulation results. In contrast to the fully inertial case, the enhancement in the steady state MSD and the corresponding dip in the kurtosis are significantly less pronounced, indicating the crucial role of rotational inertia in amplifying the resonance-like features observed in the system.

\section{MODEL AND METHOD}\label{sec:model}
We consider an inertial, chiral, active Brownian particle with mass $m$, which moves in two dimensions (2D). The particle is additionally confined to a 2D harmonic potential $V(r) = \frac{1}{2} k r^2$, with $k$ being the strength of the harmonic confinement. The Langevin equation of motion is given by
\begin{equation}
    m \vb*{\ddot{r}} = -\gamma \left( \vb*{\dot{r}} - v_s \vu*{n} \right) - k \vb*{r} + \sqrt{2 D_T} \vb*{\eta}(t),
    \label{eq:model_trans}
\end{equation}
where $\vb*{r}(t) = x(t) \vu*{i} + y(t) \vu*{j}$ is the position vector, $\gamma$ is the viscous drag, and 
$\vu*{n} = \cos\phi(t) \vu*{i} + \sin\phi(t) \vu*{j}$ is a unit vector which represents the orientation of the particle, with $\phi(t)$ being the orientation angle. The evolution of $\phi(t)$ is given by
\begin{equation}
    I \ddot{\phi}(t) = -\gamma_R \left[\dot{\phi}(t) - \Omega \right]+ \sqrt{2 D_R} \zeta(t). 
    \label{eq:model_orient}
\end{equation}
Here, $I$ is the moment of inertia of the particle about an axis perpendicular to the 2D plane, and $\gamma_R$ is the rotational friction coefficient~\cite{lowen2020inertial}. The term $\Omega$ represents the chirality of the particle, which is modelled as a torque-like effect on the orientation of the particle. 

We now discuss the physical quantities of interest. The mean-square displacement $\msd$ of the particle is given by
\begin{equation}
    \msd = \left\langle |\vb*{r}(t) - \vb*{r}_0(t) |^2 \right\rangle.
\end{equation}
Similarly, the kurtosis of position gives a quantitative estimate of how much the position distribution deviates from the Gaussian behavior, and is defined, in the steady state, as
\begin{equation}
    \kappa_r = \lim_{t\to\infty}\frac{\langle r^4(t) \rangle}{\langle r^2(t) \rangle^2}.
    \label{eq:kurtosis_def}
\end{equation}
Since we are interested in the steady-state behavior, the limit $t\to \infty$ is taken into consideration. For a Gaussian distribution $P(x,~y)$ in two dimensions, it can be shown that $\kappa_r = 2$~\cite{pattanayak2024impact}.

The simulation of the dynamics is carried out by integrating Eqs.~\eqref{eq:model_trans} and \eqref{eq:model_orient} using the modified Euler scheme with a timestep of $10^{-3}$. Each trajectory is integrated over $10^4$ timesteps, and the process is repeated for $10^5$ independent realizations to compute ensemble averages of the physical quantities.

\section{RESULTS AND DISCUSSION}\label{sec:result}
\begin{figure}
    \centering
    \includegraphics[width=\linewidth]{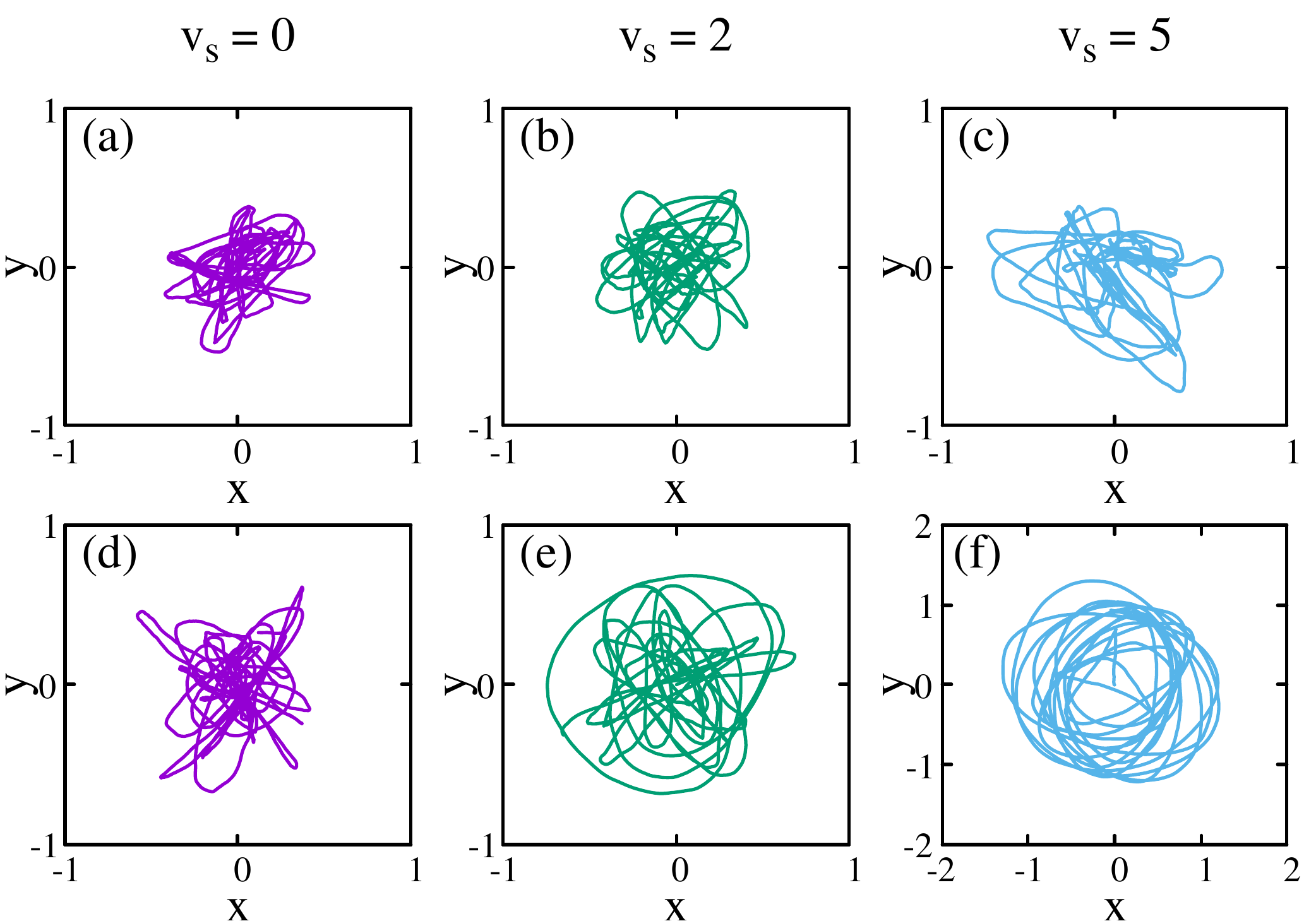}
    \caption{The simulated particle trajectory for $v_s = 0$ in (a) and (d), for $v_s = 2$ in (b) and (e), and for $v_s = 5$ in (c) and (f). In the top row, $\Omega = 0$ is taken, and in the bottom row, $\Omega = 5$ is taken. Other common parameters are $\omega_0 = 5$, and $m = I = \gamma = \gamma_R = D_T = D_R = 1$.}
    \label{fig:traj_with_vs}
\end{figure}
To characterize the nature of the particle dynamics, we first inspect the simulated trajectories for different values of the self-propulsion speed $v_s$, as shown in Fig.~\ref{fig:traj_with_vs}. In the absence of chirality [see Figs.~\ref{fig:traj_with_vs}(a)-(c)], the particle exhibits random motion, which becomes increasingly directed as $v_s$ increases, as shown in Fig.~\ref{fig:traj_with_vs}(c). In the presence of finite chirality, the trajectory appears completely random in the absence of self-propulsion or activity [see Fig.~\ref{fig:traj_with_vs}(d) for $v_s = 0$]. However, as the value of $v_s$ increases, a circular motion emerges in the trajectory, as shown in Fig.~\ref{fig:traj_with_vs}(e) and (f). Thus, the influence of chirality is stronger at higher activity. Therefore, throughout the rest of the paper, we fix $v_s = 5$ unless stated otherwise, so that the effect of chirality remains apparent.

\begin{figure}
    \centering
    \includegraphics[width=\linewidth]{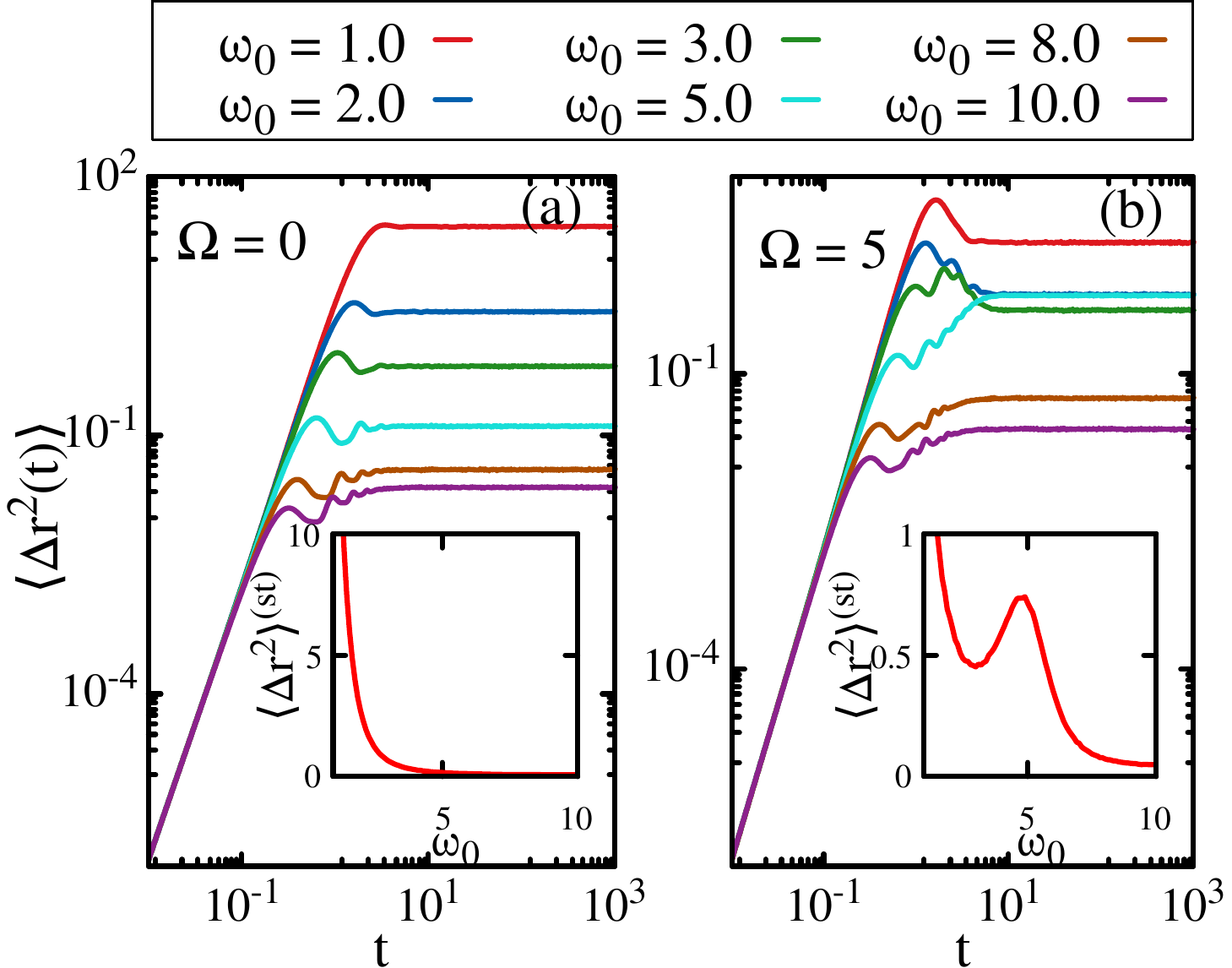}
    \caption{Plot of $\msd$ as a function of time for different values of $\omega_0$ is shown in (a) for $\Omega = 0$ and in (b) for $\Omega = 5$. The corresponding steady state value of $\msd \left(= \ev{\Delta r^2}^{(st)}\right)$ as a function of $\omega_0$ is shown in the inset of each figures. The other common parameters are $v_s = 5$, and $m = I = \gamma = \gamma_R = D_T = D_R = 1$.}
    \label{fig:msd}
\end{figure}

In Fig.~\ref{fig:msd}, we have plotted the simulated mean-square displacement $\msd$ as a function of time for different values of $\omega_0$ and for $\Omega=0$ in Fig.~\ref{fig:msd}(a) and for $\Omega=5$ in Fig.~\ref{fig:msd}(b). The $\msd$ shows short-time ballistic growth and long-time non-diffusive behavior. At intermediate times, the msd exhibits oscillation that arises from the combined effects of chirality and inertia \cite{}, which becomes more pronounced as $\omega_0$ increases. In the absence of chirality [Fig.~\ref{fig:msd}(a)], the steady-state value of $\msd$ decreases with increasing $\omega_0$, reflecting the role of confinement. The inset of Fig.~\ref{fig:msd}(a) displays the steady-state value of $\msd \left( = \ev{\Delta r^2}^{(st)} \right)$ as a function of $\omega_0$, which confirms that it decreases monotonically.
However, when finite chirality is present in the dynamics [see Fig.~\ref{fig:msd}(b)], $\ev{\Delta r^2}^{(st)}$ initially decreases with increasing $\omega_0$, then increases and reaches a maximum value when $\omega_0$ and $\Omega$ are of the same order. For larger $\omega_0$, $\ev{\Delta r^2}^{(st)}$ eventually decays to zero. The same behavior is also evident from the inset of Fig.~\ref{fig:msd}(b), where $\ev{\Delta r^2}^{(st)}$ is plotted as a function of $\omega_0$.

\begin{figure}
    \centering
    \includegraphics[width=0.95\linewidth]{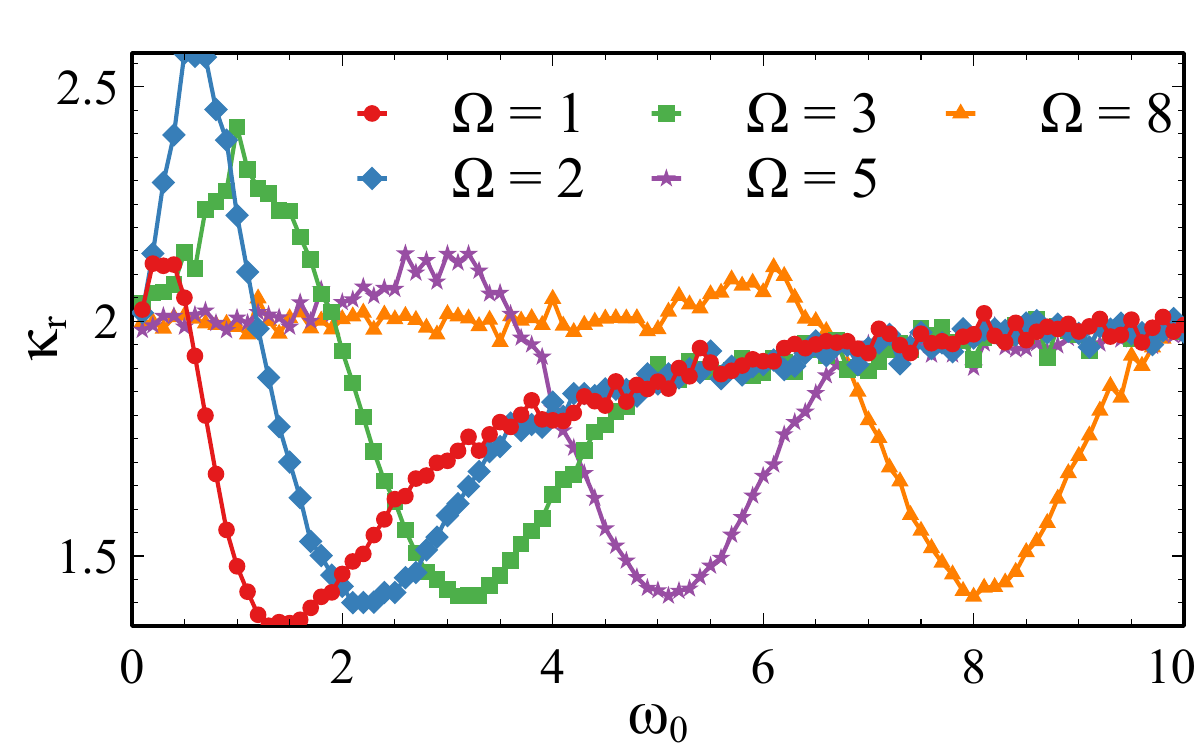}
    \caption{Kurtosis $\kappa_r$ as a function of $\omega_0$ for different values of $\Omega$. Other common parameters are: $v_s = 5$, and $m = I = \gamma = \gamma_R = D_T = D_R = 1$.}
    \label{fig:kurtosis}
\end{figure}

\begin{figure}
    \centering
    \includegraphics[width=0.8\linewidth]{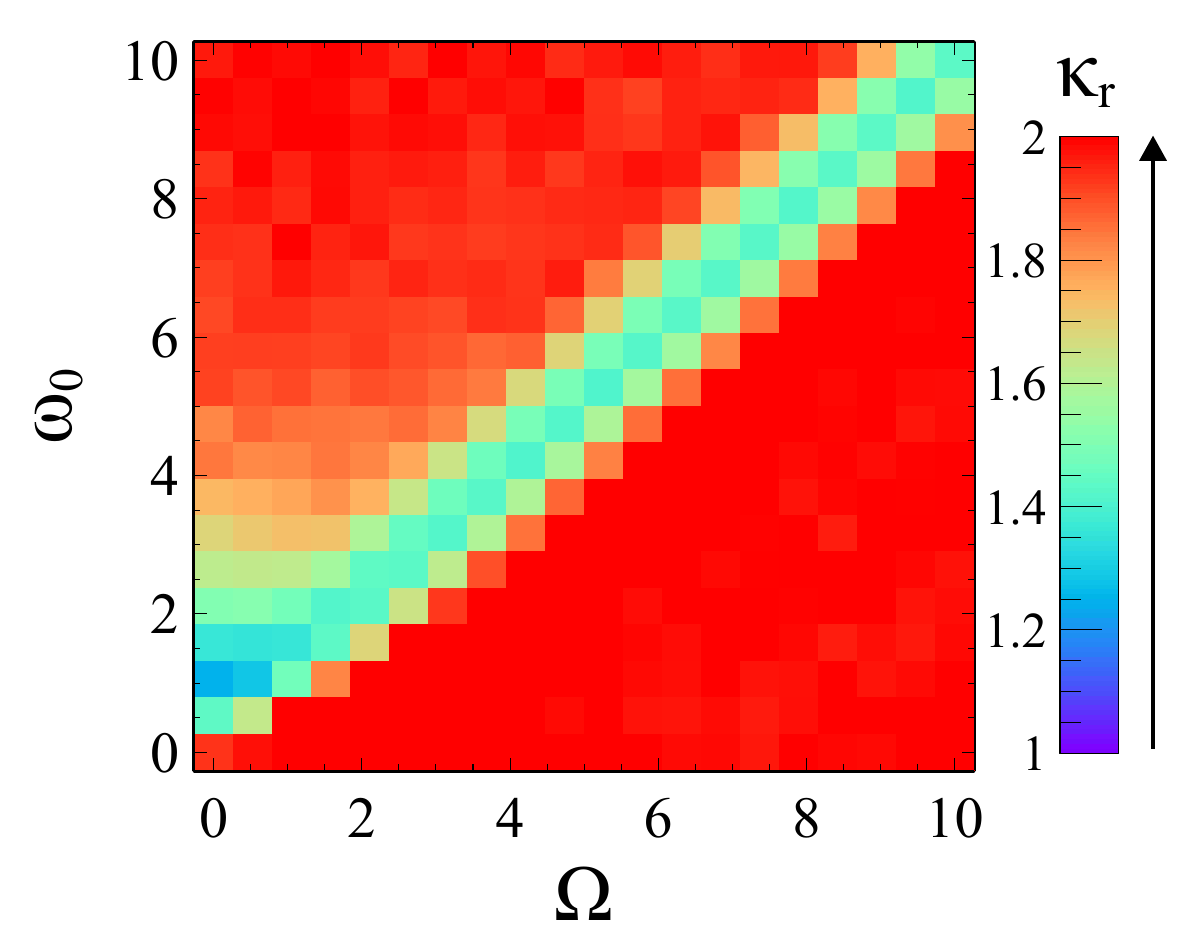}
    \caption{2D parametric plot of kurtosis $\kappa_r$ as a function of $\Omega$ and $\omega_0$. Other common parameters are: $v_s = 5$, and $m = I = \gamma = \gamma_R = D_T = D_R = 1$.}
    \label{fig:kurtosis3d}
\end{figure}

To further investigate this behavior, in Fig.~\ref{fig:kurtosis}, we show kurtosis $\kappa_r$, as defined in Eq.~\eqref{eq:kurtosis_def}, as a function of $\omega_0$ for different values of $\Omega$. For lower $\omega_0$ values, kurtosis takes the value $\kappa_r = 2$, corresponding to a Gaussian distribution \cite{}. As $\omega_0$ increases, $\kappa_r$ first increases, then decreases to a dip, and subsequently increases again before eventually saturating at $\kappa_r = 2$. The initial increase of $\kappa_r$ with $\omega_0$ becomes weaker for higher values of $\Omega$. Interestingly, the dip occurs at $\omega_0 \approx \Omega$, i.e., when the chiral frequency and the harmonic frequency are of the same order in magnitude. The same behaviour is also evident from Fig.~\ref{fig:kurtosis3d}, where $\kappa_r$ is plotted as a function of $\omega_0$ and $\Omega$ in $\Omega-\omega_0$ parametric plane. For larger values of $\Omega$ and $\omega_0$, $\kappa_r$ deviates from its Gaussian value ($\kappa_r = 2$) along the diagonal ($\omega_0 = \Omega$) of the $\Omega-\omega_0$ parameter plane. However, for smaller values of $\Omega$, the deviation is not strictly diagonal, and in the limit $\Omega \to 0$, one observes a transition from Gaussian to non-Gaussian and then back to Gaussian behavior as $\omega_0$ increases, consistent with the results reported in Ref.~\cite{kanaya2020steady}.

\begin{figure}
    \centering
    \includegraphics[width=\linewidth]{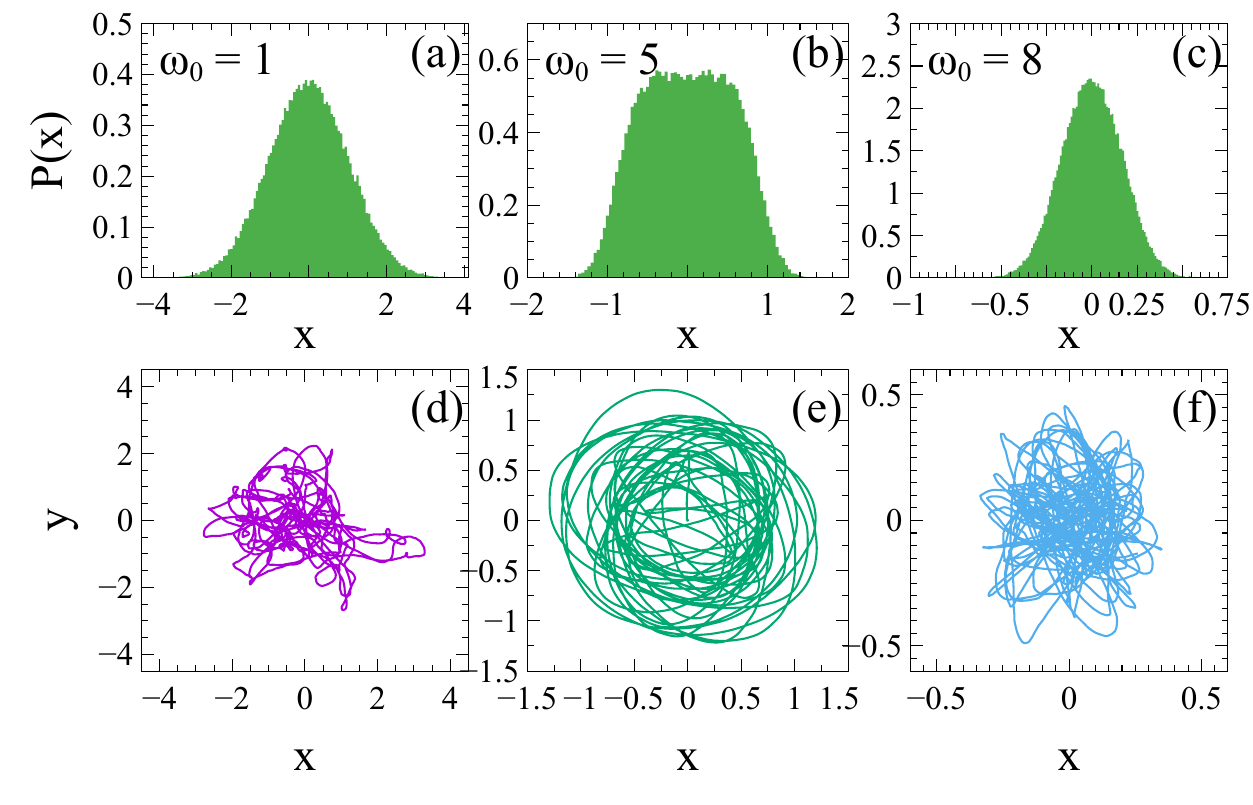}
    \caption{The marginal probability distribution $P(x)$ is plotted for $\omega_0 = 1$ in (a), for $\omega_0 = 5$ in (b) and for $\omega_0 = 8$ in (c). Corresponding trajectories are plotted in (d)-(f). Other common parameters are $\Omega = v_s = 5$, and $m = I = \gamma = \gamma_R = D_T = D_R = 1$.}
    \label{fig:Px}
\end{figure}
\begin{figure*}
    \centering
    \includegraphics[width=1\linewidth]{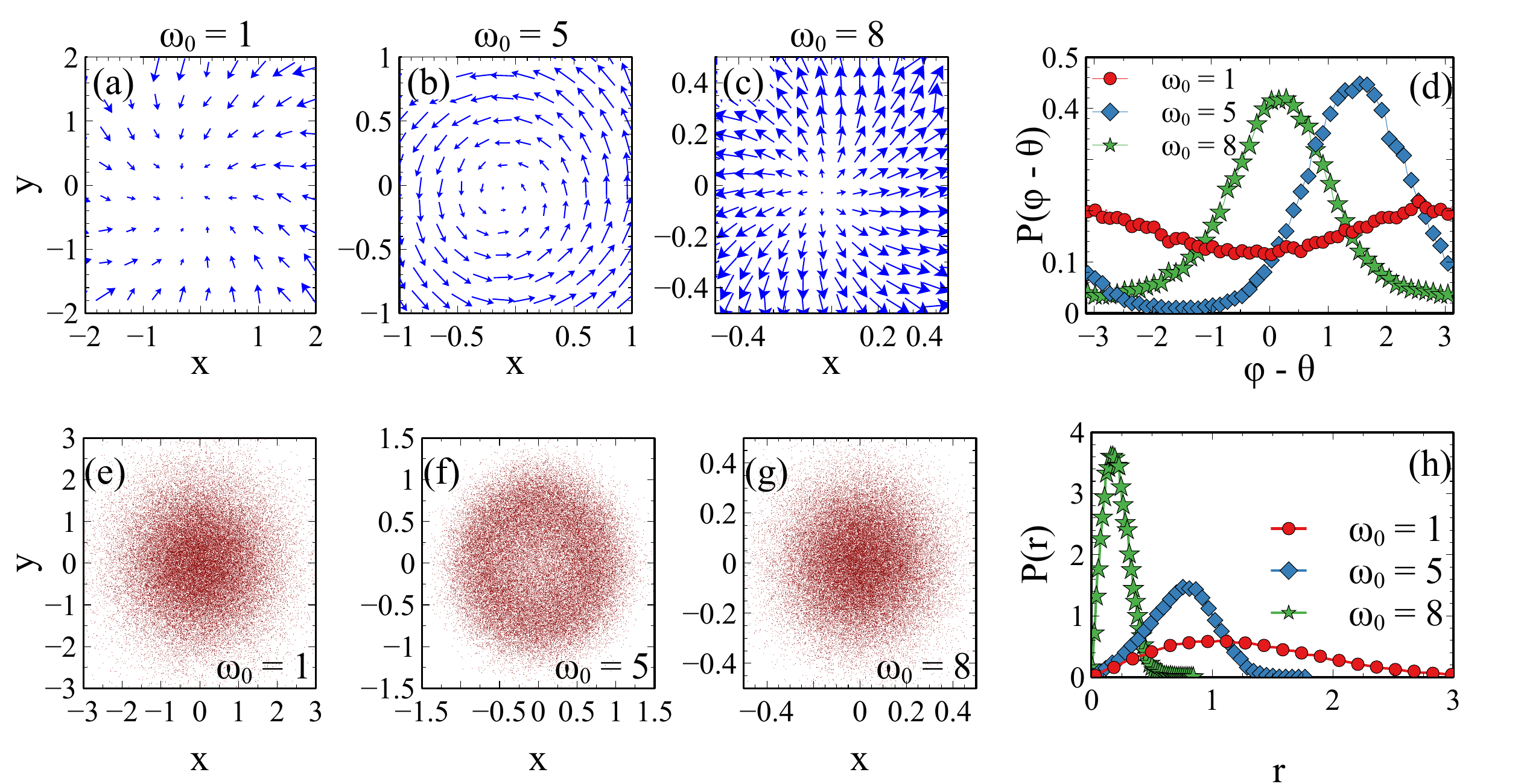}
    \caption{The simulated orientation profile $p_\phi(\vb*{r})$ is plotted in (a) for $\omega_0 = 1$, in (b) for $\omega_0 = 5$, and in (c) for $\omega_0 = 8$. In (d), the distribution $P(\phi - \theta)$ is shown. The particle distribution corresponding to the orientation profile (a)-(c) are shown in (e)-(g), respectively. The corresponding radial distribution is shown in (h). Other common parameters are $\Omega = v_s = 5$, and $m = I = \gamma = \gamma_R = D_T = D_R = 1$.}
    \label{fig:phi_and_r}
\end{figure*}

In Fig.~\ref{fig:Px}, we show the plot of marginal probability distribution $P(x)$ as a function of $x$ for different values of $\omega_0$ in Fig.~\ref{fig:Px} (a)-(c) and for $\Omega = 5$. The corresponding particle trajectories are shown in Fig.~\ref{fig:Px} (d)-(f), respectively. It can be seen that when $\omega_0$ is small [see Fig.~\ref{fig:Px}(a)], the probability distribution has a Gaussian form. 
This indicates that the particle is mostly concentrated towards the center of the potential, as evident from the particle trajectory in Fig.~\ref{fig:Px}(d). 
However, when $\omega_0$ and $\Omega$ are of the same order [see Fig.~\ref{fig:Px}(b) for $\omega_0 = \Omega = 5$], the position distribution transforms from a Gaussian to a platykurtic distribution with $\kappa_r < 2$. 
The distribution function $P(x)$ exhibits characteristically short tails and a flat region near the potential minimum. This nearly uniform distribution around the center indicates a persistent rotational motion of the particle around the potential minimum, as clearly noticeable from Fig.~\ref{fig:Px}(e). 
When the value of $\omega_0$ becomes higher than $\Omega$, the confinement plays a dominant role, as a result of which the distribution again becomes Gaussian, and the rotational trajectories disappear (see Figs.~\ref{fig:Px}(c) and ~\ref{fig:Px}(f)). 
Furthermore, if $P(\vb*{r}, \phi)$ represents the marginal probability density of the particle at position $\vb*{r}$ having an orientation $\vu*{n}(\phi)$, then, the orientation profile $\profile$ can be defined as
\begin{equation}
    \profile = \frac{\int\limits_0^{2\pi} \vu*{n}(\phi) P(\vb*{r}, \phi) \dd{\phi}}{\int\limits_0^{2\pi} P(\vb*{r}, \phi) \dd{\phi}}.
\end{equation}
In Figs.~\ref{fig:phi_and_r} (a)-(c), we show $\profile$ for different values of $\omega_0$. When $\omega_0 < \Omega$ [Fig.~\ref{fig:phi_and_r}(a)], $\profile$ spirals toward the origin, implying that the particle at most of the positions has an orientation $\phi$ nearly opposite to the radial direction $\theta$. To confirm this, in Fig.~\ref{fig:phi_and_r}(d), we have plotted the distribution $P(\phi - \theta)$ of the angle between orientation and radial direction. It can be observed that in Fig.~\ref{fig:phi_and_r}(d) for $\omega_0 = 1$ (red circles), the peak of $P(\phi - \theta)$ occurs at $\phi - \theta \approx \pi$, indicating that the particle orientations are predominantly opposite to the radial direction. Consequently, the particle distribution [Fig.\ref{fig:phi_and_r}(e)] exhibits a Gaussian form. 
Similarly, when $\omega_0 > \Omega$ [Fig.~\ref{fig:phi_and_r}(c)], $\profile$ spirals away from the origin, implying that the orientation aligns with the radial direction. This is also evident from Fig.~\ref{fig:phi_and_r}(d), when $\omega_0 > \Omega$ ($\omega_0=8$), the peak of $P(\phi - \theta)$ occurs at $\phi - \theta = 0$. As a result, the particle distribution is again Gaussian [Fig.~\ref{fig:phi_and_r}(g)]. 
Interestingly, when $\omega_0 = \Omega$ ($=5$), [Fig.~\ref{fig:phi_and_r}(b)], the orientation becomes perpendicular to the radial direction, leading to consistent rotations in $\profile$. This is confirmed in Fig.~\ref{fig:phi_and_r}(d), where for $\omega_0 = 5$, the peak of $P(\phi - \theta)$ is around $\phi - \theta \approx \pi/2$. Consequently, the particle distribution develops a ring-like structure, as shown in Fig.~\ref{fig:phi_and_r}(f).
Additionally, in Fig.~\ref{fig:phi_and_r}(h), we plot the radial position distribution $P(r)$ for the parameters used in Fig.~\ref{fig:phi_and_r}(a)-(c). For a fixed $\omega_0$ value, the distribution initially increases from zero, reaches a maximum, and then decreases back to zero as $r$ increases. As $\omega_0$ increases, the peak of the distribution increases and shifts towards lower values of $r$, while the width narrows, indicating that the particle becomes confined to a smaller region of the 2D plane, as expected. Additionally, for $\omega_0 = \Omega$($=5$), the initial increase of $P(r)$ is more linear than in the other two cases, possibly due to the ring-like distribution observed in Fig.~\ref{fig:phi_and_r}(f).

Next, we focus on the case when $I = 0$, which corresponds to the rotational overdamped limit. In such a model setup, the mass of the particle is concentrated near its center of mass, resulting in a negligible moment of inertia.
In this limit, Eq.~\eqref{eq:model_orient} becomes
\begin{equation}
    \dot{\phi}(t) = \Omega + \sqrt{\frac{2}{\tau_R}} \zeta(t),
    \label{eq:model_orient_od}
\end{equation}
where $\tau_R = {\gamma_R^2}/{D_R}$ is the persistence time induced by the rotational diffusion~\cite{Caprini2023chiral}. The Eq.~\eqref{eq:model_trans} can be written in terms of the velocity of the particle $\vb*{v} (= \vb*{\dot{r}})$ as
\begin{equation}
    \vb*{\dot{v}} = -\frac{1}{\tau_m} \left( \vb*{v} - v_s \vu*{n} \right) - \omega_0^2 \vb*{r} + \frac{\sqrt{2 D_T}}{m} \vb*{\eta}(t),
    \label{eq:model_trans_v}
\end{equation}
where $\tau_m = \frac{m}{\gamma}$ is the inertial timescale and $\omega_0 = \sqrt{\frac{k}{m}}$ is the harmonic frequency. The Fokker-Planck equation for the probability distribution $P = P(\vb*{r}, \vb*{v}, \phi; t)$ corresponding to Eqs.~\eqref{eq:model_orient_od} and \eqref{eq:model_trans_v} can be written as
\begin{equation}
\begin{split}
    \pdv{P}{t} &= - \div(\vb*{v}P) + \frac{1}{\tau_m} \vdiv\left[ (\vb*{v} - v_s \vu*{n})P \right] + \omega_0^2 \vdiv(\vb*{r}P) \\
    & + \frac{D_T}{m^2} \vlaplacian{P} - \pdv{\phi}(\Omega P) + \frac{1}{\tau_R} \pdv[2]{P}{\phi},
\end{split}
\label{eq:FPE_def}
\end{equation}
where $\grad$ and $\vgrad$ are the partial differential operators in position and velocity space, respectively. Taking Laplace transform of Eq.~\eqref{eq:FPE_def} with respect to $t$, we get
\begin{equation}
\begin{split}
    s \Tilde{P} - P_0 &= -\div(v\Tilde{P}) + \frac{1}{\tau_m}\vdiv\left[ (\vb*{v} - v_s \vu*{n}) \Tilde{P} \right] \\
    & + \omega_0^2 \vdiv\left(r \Tilde{P}\right) + \frac{D_T}{m^2} \vlaplacian\Tilde{P} \\
    &- \pdv{\phi}(\Omega \Tilde{P}) + \frac{1}{\tau_R} \pdv[2]{\Tilde{P}}{\phi}.
\end{split}
\label{eq:FPE_LT}
\end{equation}
Here, $\Tilde{P} = \Tilde{P}(\vb*{r}, \vb*{v}, \phi; s) = \int_0^\infty\; \dd{t} e^{-st} P(\vb*{r}, \vb*{v}, \phi; t)$ is the Laplace transform of the probability distribution $P$. The initial distribution $P_0~=~P(\vb*{r}, \vb*{v}, \phi; 0)$ can be chosen, without the loss of generality, as
\begin{equation}
    P(\vb*{r}, \vb*{v}, \phi; 0) = \delta(\vb*{r}) \delta(\vb*{v})\delta(\phi - \phi_0),
\end{equation}

where $\phi_0$ is the initial orientation of the particle. Even though it is not possible to solve Eq.~\eqref{eq:FPE_def} due to the nonlinear term $v_s\vu*{n}$, one can evaluate all the moments of an arbitrary dynamical variable $\Psi = \Psi(\vb*{r}, \vb*{v}, \phi)$ as discussed in Ref.~\cite{patel2023exact}. Multiplying Eq.~\eqref{eq:FPE_LT} by $\Psi$ and integrating over $\vb*{r}$, $\vb*{v}$ and $\phi$, we get
\begin{equation}
\begin{split}
    s \ev{\Psi}_s &- \ev{\Psi}_0 = \ev{\vb*{v}\vdot \grad \Psi}_s - \frac{1}{\tau_m} \ev{\vb*{v} \vdot\vgrad\Psi}_s \\
    &+ \frac{v_s}{\tau_m} \ev{\vu*{n} \vdot \vgrad\Psi}_s - \omega_0^2 \ev{\vb*{r} \vdot \vgrad\Psi}_s \\
    & + \frac{D_T}{m^2} \ev{\vlaplacian \Psi}_s + \Omega \ev{\pdv{\Psi}{\phi}}_s + \frac{1}{\tau_R} \ev{\pdv[2]{\Psi}{\phi}}_s,
\end{split}
\label{eq:moment_def}
\end{equation}
where
\begin{equation}
    \ev{\Psi}_s = \int\dd{\vb*{r}}\int\dd{\vb*{v}}\int\dd{\phi} \Psi(\vb*{r}, \vb*{v}, \phi) \Tilde{P}(\vb*{r}, \vb*{v}, \phi; s),
\end{equation}
and
\begin{equation}
    \ev{\Psi}_0 = \int\dd{\vb*{r}}\int\dd{\vb*{v}}\int\dd{\phi} \Psi(\vb*{r}, \vb*{v}, \phi) P(\vb*{r}, \vb*{v}, \phi; 0).
\end{equation}
For convenience, we define
\begin{align}
    \vb*{\rll} &= (\vb*{r}\vdot \vu*{n})\vu*{n}; \qquad \vb*{\rT} = \vu*{n} \cross (\vb*{r}\cross \vu*{n}), \\
    \text{and} \qquad 
    \vb*{\vll} &= (\vb*{v}\vdot \vu*{n})\vu*{n}; \qquad \vb*{\vT} = \vu*{n} \cross (\vb*{v}\cross \vu*{n}).
\end{align}
Here, $\vb*{\rll}$ and $\vb*{\vll}$ are respectively the components of $\vb*{r}$ and $\vb*{v}$ along the orientation of the particle, and $\vb*{\rT}$ and $\vb*{\vT}$ are those perpendicular to it.
The Eq.~\eqref{eq:moment_def} can be used to find $\ev{r^2}^{(st)}$ and $\kappa_r$. To do this, first we put $\Psi = r^2$ along with the initial conditions $\vb*{r_0} = \vb*{v_0} = 0$ in Eq.~\eqref{eq:moment_def}. This will give 
\begin{equation}
    s \ev{r^2}_s = 2 \ev{\vb*{r} \vdot \vb*{v}}_s.
    \label{eq:moment_r2}
\end{equation}
In order to solve Eq.~\eqref{eq:moment_r2}, $\ev{\vb*{r} \vdot \vb*{v}}_s$ must be known. Setting $\Psi = \vb*{r} \vdot \vb*{v}$ in Eq.~\eqref{eq:moment_def}, we get
\begin{equation}
    \left(s + \frac{1}{\tau_m}\right) \ev{\vb*{r} \vdot \vb*{v}}_s = \ev{v^2}_s + \frac{v_s}{\tau_m} \ev{\rll}_s - \omega_0^2 \ev{r^2}_s.
    \label{eq:moment_rv}
\end{equation}
Similarly, when $\Psi = v^2$, we get
\begin{equation}
    \left(s+\frac{2}{\tau_m}\right) \ev{v^2}_s = \frac{2 v_s}{\tau_m} \ev{\vll}_s - 2 \omega_0^2 \ev{\vb*{r} \vdot \vb*{v}}_s + \frac{4 D_T}{s m^2}.
    \label{eq:moment_v2}
\end{equation}
\begin{figure}
    \centering
    \includegraphics[width=\linewidth]{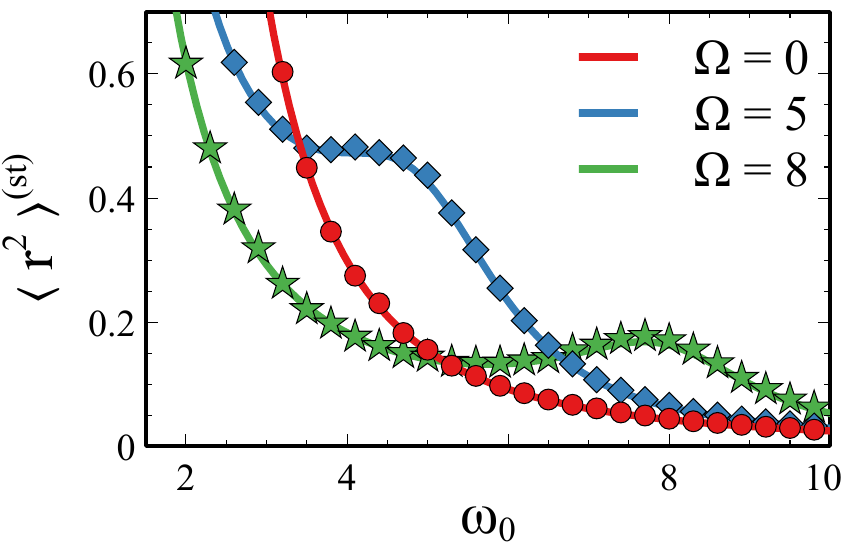}
    \caption{Plot of $\ev{r^2}^{(st)}$[Eq.~\eqref{eq:msd_st}] as a function of $\omega_0$ for different values of $\Omega$. Solid lines represent the analytical result, and points represent the simulation. Other common parameters are $v_s = 5$ and $\tau_m = \tau_R = m = D_T = 1$.}
    \label{fig:msdst_no_RI}
\end{figure}
Proceeding in a similar way, we will arrive at the following set of equations
\begin{align}
    &\left( s + \frac{1}{\tau_R} \right)\ev{\rll}_s = \ev{\vll}_s - \Omega \ev{\rT}_s \label{eq:moment_rdotn} \\
    &\left( s + \frac{1}{\tau_R}\right)\ev{\rT}_s = \ev{\vT}_s + \Omega \ev{\rll}_s \label{eq:moment_rcrossn} \\
    &\left( s + \frac{1}{\tau_m} + \frac{1}{\tau_R} \right)\ev{\vll}_s = \frac{v_s}{s\tau_m} - \omega_0^2 \ev{\rll}_s - \Omega \ev{\vT}_s \label{eq:moment_vdotn}\\
    &\left( s + \frac{1}{\tau_m} + \frac{1}{\tau_R} \right)\ev{\vT}_s = -\omega_0^2 \ev{\rT}_s + \Omega \ev{\vll}_s. \label{eq:moment_vcrossn}
\end{align}
Solving Eqs.~\eqref{eq:moment_r2}-\eqref{eq:moment_vcrossn}, one can obtain $\ev{r^2}_s$. The expression for $\ev{r^2}_s$ is 
given by 
\begin{widetext}
\begin{equation}
\begin{split}
    \ev{r^2}_s &= \frac{8 D_T \tau_m^2}{m^2 \Lambda} + \frac{8v_s^2 \tau_R\left( 2 + s \tau_R \right)}{\left( 4 + 4s \tau_R + s^2\tau_R^2 + 4 \tau_R^2\Omega^2 \right) \Lambda} \\
    &- \frac{2 v_s^2 \left\{\tau _m \left[\tau _R \left(s^3 \tau _R^2+4 s^2 \tau _R+\omega _0^2 \tau _R \left(s \tau _R+2\right)-5 s \Omega ^2 \tau _R^2-6 \Omega ^2 \tau _R+5
   s\right)+2\right]+\tau _R \left[\tau _R \left(s \left(s \tau _R+3\right)-2 \Omega ^2 \tau _R\right)+2\right]\right\}}{\alpha _1 s \left[\left(2 \alpha_2 - s\right)^2+4 \Omega ^2\right] \tau _m^3 \tau _R^3 \left[2 \omega _0^2 \left(\alpha _2 \alpha _3-\Omega
   ^2\right)+\left(\alpha _2^2+\Omega ^2\right) \left(\alpha _3^2+\Omega ^2\right)+\omega _0^4\right]},
\end{split}
\label{eq:r2s_final}
\end{equation}
where
\begin{align}
    \Lambda &= s \left(s \tau _m+1\right) \left(s \left(s \tau _m+2\right)+4 \omega _0^2 \tau _m\right), \\
    \alpha_1 &= s + \frac{1}{\tau_m}, \ 
    \alpha_2 = s + \frac{1}{\tau_R}, \  \text{and} \ 
    \alpha_3 = s + \frac{1}{\tau_m} + \frac{1}{\tau_R}. \\
\end{align} 
To obtain the steady state value of $\ev{r^2(t)} \left( = \ev{r^2}^{(st)}\right)$ from $\ev{r^2}_s$, we have
\begin{equation}
    \ev{r^2}^{(st)} = \lim_{t \to \infty} \ev{r^2(t)} = \lim_{s\to 0} s \ev{r^2}_s.
    \label{eq:final_value_theorem}
\end{equation}
Substituting $\ev{r^2}_s$ from Eq.~\eqref{eq:r2s_final} into Eq.~\eqref{eq:final_value_theorem}, we get
\begin{equation}
    \ev{r^2}^{(st)} = \frac{2 D_T \tau_m}{m^2 \omega_0^2} + \frac{\tau _R v_s^2 \left[\omega _0^2 \tau _m \tau _R^2 \left(\tau _m+\tau _R\right)+\Omega ^2 \tau _m^2 \tau _R^2+2 \tau _m \tau _R+\tau _m^2+\tau _R^2\right]}{\omega _0^2
   \tau _m \left\{2 \omega _0^2 \tau _m \tau _R^2 \left(-\Omega ^2 \tau _m \tau _R^2+\tau _m+\tau _R\right)+\omega _0^4 \tau _m^2 \tau _R^4+\left(\Omega ^2 \tau
   _R^2+1\right) \left[\Omega ^2 \tau _m^2 \tau _R^2+\left(\tau _m+\tau _R\right)^2\right]\right\}}.
   \label{eq:msd_st}
\end{equation}
\end{widetext}
The $\ev{r^2}^{(st)}$ contains two terms. The First term of $\ev{r^2}^{(st)}$ is the steady state mean square displacement of a passive Brownian particle. The second term of Eq.~\eqref{eq:msd_st} is the contribution due to the activity, and it approaches zero in the $v_s \to 0$ limit. 
In Fig.~\ref{fig:msdst_no_RI}, we have plotted $\ev{r^2}^{(st)}$ as a function of $\omega_0$ for different values of $\Omega$. When $\Omega = 0$, that is, in the absence of chirality, the $\ev{r^2}^{(st)}$ exhibits a monotonic decay with $\omega_0$. However, when $\Omega$ is finite, $\ev{r^2}^{(st)}$ initially decreases, and then increases and exhibits a peak near $\omega_0 \approx \Omega$, and then finally decreases with $\omega_0$ and approaches zero. The presence of the peak is similar to the case of dynamics with finite rotational inertia, as shown in the inset of Fig.~\ref{fig:msd}(b). However, the peak is more pronounced in the case when $I \neq 0$. The value of $\omega_0 \left( = \omega_{0,\text{peak}} \right)$ can be found by analyzing Eq.~\eqref{fig:msdst_no_RI}. Since the numerator of the second term of Eq.~\eqref{fig:msdst_no_RI} is an increasing function of $\omega_0$, the peak appears when the denominator of Eq.~\eqref{fig:msdst_no_RI} is minimum. Taking the derivative of the denominator of the second term in Eq.~\eqref{fig:msdst_no_RI}, and equating to zero, we get
\begin{equation}
    \omega_{0, \text{peak}} = \sqrt{\Omega^2 - \frac{1}{\tau_R \tau_{eff}}},
\end{equation}
where $\tau_{eff} = \frac{\tau_m \tau_R}{\tau_m + \tau_R}$ is the effective timescale due to inertia and rotational diffusion. For a definite value of $\tau_R$ and $\tau_m$, we can see that $\omega_{0, \text{peak}} \approx \Omega$ for higher values of $\Omega$.

Similarly, to calculate $\kappa_r$, one can follow the procedure discussed above. The detailed calculation is given in appendix~\ref{sec:App:B}. In Fig.~\ref{fig:kr_no_RI}, we plot $\kappa_r$ as a function of $\omega_0$ for different values of $\Omega$. The behavior closely resembles that in Fig.~\ref{fig:kurtosis}, as $\kappa_r$ initially increases above the value $\kappa_r = 2$ with an increase in $\omega_0$, then decreases and exhibits a dip around $\omega_0 \approx \Omega$. However, the initial rise is significantly less pronounced than in the presence of rotational inertia. In addition, the minimum occurs at a higher value of $\kappa_r$ compared to Fig.~\ref{fig:kurtosis}, where rotational inertia is present.
\begin{figure}
    \centering
    \includegraphics[width=\linewidth]{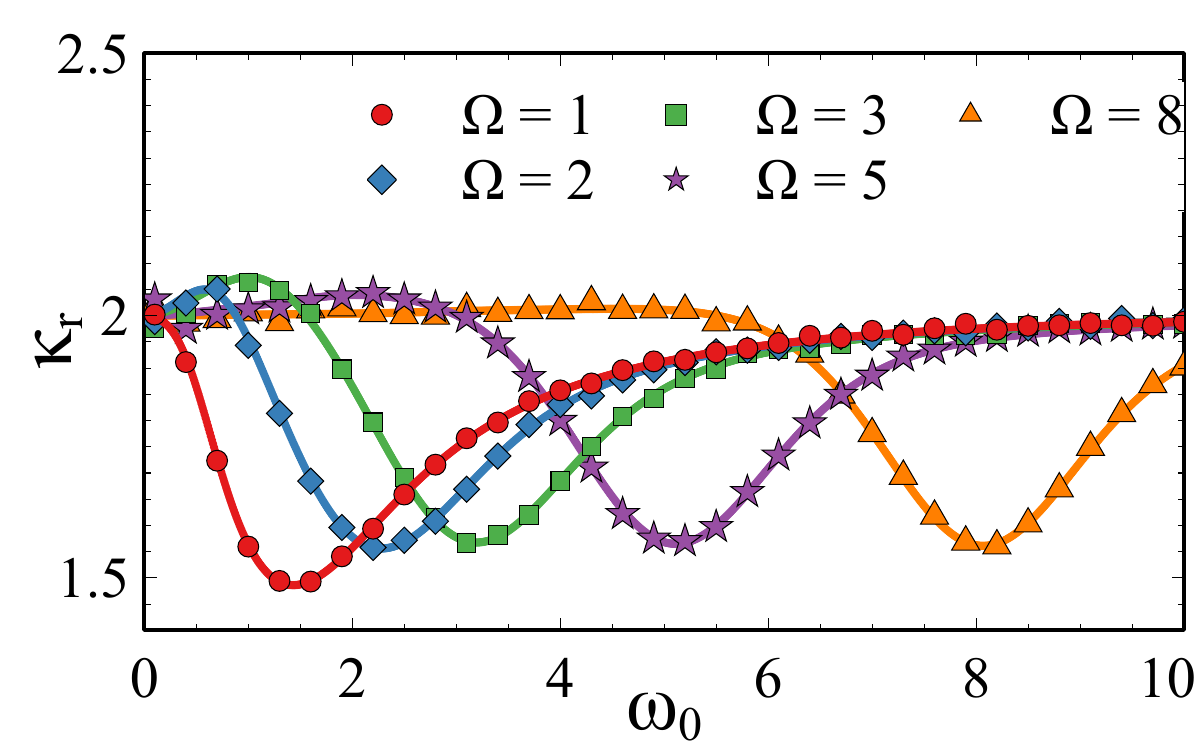}
    \caption{The plot of $\kappa_r$ for the rotationally overdamped case as a function of $\omega_0$ for different values of $\Omega$. Solid lines represent the analytical result, and points represent the simulation. Other common parameters are $v_s = 5$ and $\tau_m = \tau_R = m = D_T = 1$.}
    \label{fig:kr_no_RI}
\end{figure}

\section{CONCLUSIONS}\label{sec:summary}
In conclusion, we have investigated the transport behavior of an inertial chiral active Brownian particle confined in a harmonic potential. Using numerical simulations and an analytical approach, we analyzed the particle trajectories, mean-square displacement, and steady-state kurtosis. In the absence of chirality, the trajectories exhibit persistent motion, whereas in the presence of chirality, they become circular. In addition, to analyze the diffusive behavior, we have simulated the MSD of the particle. The MSD shows ballistic behavior in the short-time limit and diffusive behavior in the long-time limit. However, it exhibits oscillations on the intermediate time scale, which increase as the value of the harmonic frequency increases. In the absence of chirality, the steady-state MSD monotonically decays with increasing harmonic frequency. Interestingly, in the presence of chirality, the steady-state MSD shows a maximum when the harmonic and chiral frequencies become comparable. This feature is corroborated by steady-state kurtosis, which shows a pronounced dip when both the chiral and harmonic frequencies match. Correspondingly, the position distribution deviates from Gaussian behavior and becomes platykurtic with a flat region near the potential minimum. In this regime, the particle tends to move towards the edge of the potential and align perpendicular to the radial direction. 

To further elucidate this behavior analytically, we derived expressions for the steady-state MSD and kurtosis in the limit where the orientational dynamics is overdamped, while retaining translational inertia. The analytical predictions are in good agreement with the simulation results. Although the same qualitative features persist in this limit, the peak in the steady-state MSD and the suppression in the kurtosis are less pronounced, highlighting the role of rotational inertia in amplifying the observed resonance-like response.

\section{Acknowledgement}
MS and SD acknowledge the SERB-SURE grant (SUR/2022/000377) and the CRG grant (CRG/2023/002026) from DST, Govt. of India for financial support.

\appendix
\begin{widetext}

\section{Calculation of $\kappa_r$ for the rotationally overdamped case}\label{sec:App:B}
In order to evaluate $\kappa_r$, we need the steady state value of the fourth moment of position $\ev{r^4}^{(st)}$. 
For this, the following equations are obtained from Eq.~\eqref{eq:moment_def}.
\allowdisplaybreaks
\begin{align}
    s \ev{r^4}_s &= 4 \ev{r^2(\vb*{r}\vdot \vb*{v})}_s 
    \label{eq:r4_system_start}\\
    \left( s + \frac{1}{\tau_m} \right) \ev{r^2 (\vb*{r}\vdot \vb*{v}}_s &= \ev{r^2v^2}_s + 2 \ev{(\vb*{r}\vdot \vb*{v})^2}_s - \omega_0^2 \ev{r^4}_s + \frac{v_s}{\tau_m} \ev{r^2 \rll} \\
    \left( s + \frac{2}{\tau_m} \right) \ev{r^2 v^2}_s &= 2 \ev{v^2 (\vb*{r}\vdot \vb*{v})}_s + \frac{4 D_T}{m^2} \ev{r^2}_s - 2 \omega_0^2 \ev{r^2(\vb*{r}\vdot \vb*{v})}_s + \frac{2 v_s}{\tau_m} \ev{r^2 \vll}_s\\
    \left( s + \frac{2}{\tau_m} \right) \ev{(\vb*{r} \vdot \vb*{v})^2}_s &= \frac{2 D_T }{m^2} \ev{r^2}_s + 2 \ev{v^2 (\vb*{r} \vdot \vb*{v})}_s - 2 \omega_0^2 \ev{r^2 (\vb*{r} \vdot \vb*{v})}_s + \frac{2 v_s}{\tau_m} \ev{(\vb*{r} \vdot \vb*{v}) \rll}\\
    \left( s + \frac{3}{\tau_m} \right) \ev{v^2(\vb*{r} \vdot \vb*{v})}_s &= \ev{v^4}_s + \frac{8 D_T}{m^2} \ev{\vb*{r} \vdot \vb*{v}}_s - \omega_0^2 \ev{r^2v^2}_s - 2\omega_0^2 \ev{(\vb*{r} \vdot \vb*{v})^2}_s + \frac{v_s}{\tau_m} \ev{v^2 \rll}_s + \frac{2 v_s}{\tau_m} \ev{(\vb*{r} \vdot \vb*{v}) \vll}_s\\
    \left( s + \frac{1}{\tau_R} \right) \ev{r^2 \rll}_s &= -\Omega \ev{r^2 \rT}_s + \ev{r^2\vll}_s + 2 \ev{(\vb*{r} \vdot \vb*{v}) \rll}_s\\
    \left( s + \frac{1}{\tau_m} + \frac{1}{\tau_R} \right) \ev{r^2 \vll}_s &= -\omega_0^2 \ev{r^2\rll}_s + \frac{v_s}{\tau_m} \ev{r^2}_s - \Omega \ev{r^2 \vT}_s + 2 \ev{(\vb*{r} \vdot \vb*{v})\vll}_s \\
    \left( s + \frac{1}{\tau_R} + \frac{2}{\tau_m} \right) \ev{v^2 \rll}_s &= \ev{v^2 \vll}_s + \frac{4 D_T}{m^2} \ev{\rll}_s - \Omega \ev{v^2 \rll}_s - 2 \omega_0^2 \ev{(\vb*{r} \vdot \vb*{v}) \rll}_s + \frac{2 v_s}{\tau_m} \ev{\rll \vll}_s\\
    \left( s + \frac{3}{\tau_m} + \frac{1}{\tau_R} \right) \ev{v^2 \vll}_s &= \frac{8 D_T}{m^2} \ev{\vll}_s - \Omega \ev{v^2 \vT}_s - \omega_0^2 \ev{v^2\rll}_s - 2 \omega_0^2\ev{(\vb*{r}\vdot \vb*{v})\vll}_s + \frac{v_s}{\tau_m}\ev{v^2}_s + \frac{2 v_s}{\tau_m} \ev{\vll^2}_s \\
    \left( s + \frac{1}{\tau_R} + \frac{1}{\tau_m} \right) \ev{(\vb*{r} \vdot \vb*{v})\rll}_s &= -\Omega \ev{(\vb*{r} \vdot \vb*{v}) \rT}_s - \omega_0^2 \ev{r^2 \rll}_s + \frac{v_s}{\tau_m} \ev{\rll^2}_s + \ev{v^2\rll}_s + \ev{(\vb*{r}\vdot \vb*{v})\vll}_s \\
    \left( s + \frac{1}{\tau_R} + \frac{2}{\tau_m} \right) \ev{(\vb*{r} \vdot \vb*{v}) \vll}_s &= \frac{2 D_T}{m^2} \ev{\rll}_s - \Omega \ev{(\vb*{r} \vdot \vb*{v}) \vT}_s + \ev{v^2\vll}_s - \omega_0^2\ev{r^2\vll}_s - \omega_0^2 \ev{(\vb*{r} \vdot \vb*{v})\rll}_s \nonumber\\
    &+ \frac{v_s}{\tau_m}\ev{\vb*{r} \vdot \vb*{v}}_s + \frac{v_s}{\tau_m} \ev{\rll \vll}_s \\
    \left( s + \frac{1}{\tau_m} + \frac{2}{\tau_R} \right)
    \ev{\rll \vll}_s &=
    \ev{\vll^2}_s 
    - \omega_0^2 \ev{\rll^2}_s
    + \frac{v_s}{\tau_m} \ev{\rll}_s
    - \Omega \ev{\rll \vT}_s
    - \Omega \ev{\vll \rT}_s
    + \frac{2}{\tau_R} \ev{\rT \vT}_s \\
    \left( s + \frac{4}{\tau_m} \right) \ev{v^4}_s &=
    \frac{16 D_T}{m^2} \ev{v^2}_s
    - 4 \omega_0^2 \ev{v^2 (\vb*{r} \vdot \vb*{v})}_s
    + \frac{4 v_s}{\tau_m} \ev{v^2 \vll}_s \\
    \left( s + \frac{2}{\tau_m} + \frac{2}{\tau_R} \right)
    \ev{\vll^2}_s &=
    \frac{2 D_T}{m^2 s}
    - 2 \omega_0^2 \ev{\rll \vll}_s
    + \frac{2 v_s}{\tau_m} \ev{\vll}_s
    - 2 \Omega \ev{\vll \vT}_s
    + \frac{2}{\tau_R} \ev{\vT^2}_s \\
    \left( s + \frac{2}{\tau_R} \right) \ev{\rll^2}_s &=
    2 \ev{\rll \vll}_s
    - 2 \Omega \ev{\rll \rT}_s
    + \frac{2}{\tau_R} \ev{\rT^2}_s \\
    \left( s + \frac{1}{\tau_R} \right) \ev{r^2 \rT}_s &=
    \Omega \ev{r^2 \rll}_s
    + \ev{r^2 \vT}_s
    + 2 \ev{(\vb*{r} \vdot \vb*{v}) \rT}_s \\
     \left( s + \frac{1}{\tau_m} + \frac{1}{\tau_R} \right)
    \ev{r^2 \vT}_s &=
    - \omega_0^2 \ev{r^2 \rT}_s
    + \Omega \ev{r^2 \vll}_s
    + 2 \ev{(\vb*{r} \vdot \vb*{v}) \vT}_s \\
    \left( s + \frac{2}{\tau_m} + \frac{1}{\tau_R} \right) 
    \ev{v^2 \rT}_s &= 
    \ev{v^2 \vll}_s + \frac{4 D_T}{m^2} \ev{\rT}_s 
    - 2 \omega_0^2 \ev{(\vb*{r}\vdot\vb*{v}) \rT}_s + \frac{2 v_s}{\tau_m} \ev{\vll \rT}_s + \Omega \ev{v^2 \rll}_s \\
    \left( s + \frac{3}{\tau_m} + \frac{1}{\tau_R} \right) 
    \ev{v^2 \vT}_s &= 
    \frac{8 D_T}{m^2} \ev{\vT}_s 
    + \Omega \ev{v^2 \vll}_s
    - \omega_0^2 \ev{v^2 \rT}_s - 2 \omega_0^2 \ev{(\vb*{r}\vdot\vb*{v}) \vT}_s + \frac{2 v_s}{\tau_m} \ev{\vll \vT}_s \\
    \left( s + \frac{1}{\tau_m} + \frac{1}{\tau_R} \right) 
    \ev{(\vb*{r}\vdot\vb*{v}) \rT}_s &= 
    - \omega_0^2 \ev{r^2 \rT}_s + \frac{v_s}{\tau_m} \ev{\rll \rT}_s + \Omega \ev{(\vb*{r}\vdot\vb*{v}) \rll}_s + \ev{v^2 \rT}_s + \ev{(\vb*{r}\vdot\vb*{v}) \vT}_s \\
    \left( s + \frac{2}{\tau_m} + \frac{1}{\tau_R} \right) 
    \ev{(\vb*{r}\vdot\vb*{v}) \vT}_s &= 
    \frac{2 D_T}{m^2} \ev{\rT}_s
    + \ev{v^2 \vT}_s + \Omega \ev{(\vb*{r}\vdot\vb*{v}) \vll}_s 
    - \omega_0^2 \ev{r^2 \vT}_s - \omega_0^2 \ev{(\vb*{r}\vdot\vb*{v}) \rT}_s 
    + \frac{v_s}{\tau_m} \ev{\rll \vT}_s \\
    \left( s + \frac{4}{\tau_R} \right) 
    \ev{\rll \rT}_s &= 
    \Omega \ev{\rll^2}_s - \Omega \ev{\rT^2}_s + \ev{\rll \vT}_s + \ev{\vll \rT}_s \\
    \left( s + \frac{1}{\tau_m} + \frac{2}{\tau_R} \right) 
    \ev{\rll \vT}_s &= 
    \ev{\vll \vT}_s - \omega_0^2 \ev{\rll \rT}_s 
    - \frac{2}{\tau_R} \ev{\vll \rT}_s 
    - \Omega \ev{\rT \vT}_s + \Omega \ev{\rll \vll}_s \\
    \left( s + \frac{1}{\tau_m} + \frac{2}{\tau_R}  \right) 
    \ev{\vll \rT}_s &= 
    \ev{\vll \vT}_s - \omega_0^2 \ev{\rll \rT}_s + \frac{v_s}{\tau_m} \ev{\rT}_s - \frac{2}{\tau_R} \ev{\rll \vT}_s - \Omega \ev{\rT \vT}_s + \Omega \ev{\rll \vll}_s \\
    \left( s + \frac{2}{\tau_m} + \frac{4}{\tau_R} \right) 
    \ev{\vll \vT}_s &= 
    - \Omega \ev{\vT^2}_s + \Omega \ev{\vll^2}_s - \omega_0^2 \ev{\rll \vT}_s - \omega_0^2 \ev{\vll \rT}_s + \frac{v_s}{\tau_m} \ev{\vT}_s \\
    \left( s + \frac{2}{\tau_R} \right) 
    \ev{\rT^2}_s &= 
    2 \ev{\rT \vT}_s + 2 \Omega \ev{\rll \rT}_s + \frac{2}{\tau_R} \ev{\rll^2}_s \\
    \left( s + \frac{2}{\tau_m} + \frac{2}{\tau_R} \right) 
    \ev{\vT^2}_s &= 
    \frac{2 D_T}{m^2 s} + 2 \Omega \ev{\vll \vT}_s - 2 \omega_0^2 \ev{\rT \vT}_s + \frac{2}{\tau_R} \ev{\vll^2}_s \\
    \left( s  + \frac{1}{\tau_m}+ \frac{2}{\tau_R} \right) 
    \ev{\rT \vT}_s &= 
    \ev{\vT^2}_s - \omega_0^2 \ev{\rT^2}_s + \Omega \ev{\rll \vT}_s + \Omega \ev{\vll \rT}_s + \frac{2}{\tau_R} \ev{\rll \vll}_s \label{eq:r4_system_end}
\end{align}

Solving Eqs.~\eqref{eq:r4_system_start}-\eqref{eq:r4_system_end}, one can evaluate $\ev{r^4}_s$. Now, the steady state value of $\ev{r^4(t)} \left( = \ev{r^4}^{(st)}\right)$ can be evaluated as
\begin{equation}
    \ev{r^4}^{(st)} = \lim_{s\to 0} s \ev{r^4}_s,
\end{equation}
which gives
\begin{equation}
\begin{aligned}
    \ev{r^4}^{(st)} =\;&
    \tau_m^{2}
    \Bigg[
    \frac{8\,D_T\,\tau_m^{2}}{m^{4}}
    +
    \frac{
    8\,D_T\, v_s^{2}\, \tau_R (
    \tau_m^{2}
    + 2\tau_m \tau_R
    + \tau_R^{2}
    + \tau_m^{2} \tau_R^{2} \Omega^{2}
    + \tau_m \tau_R^{2} (\tau_m + \tau_R)\, \omega_0^{2})
    }{
    m^{2}\Bigl[
    (1 + \tau_R^{2} \Omega^{2})(
    (\tau_m + \tau_R)^{2}
    + \tau_m^{2} \tau_R^{2} \Omega^{2})
    + 2\tau_m \tau_R^{2} (\tau_m + \tau_R - \tau_m \tau_R^{2} \Omega^{2})\, \omega_0^{2}
    + \tau_m^{2} \tau_R^{4} \omega_0^{4}
    \Bigr]
    }
    \\[10pt]
    &+ \frac{v_s^4 \tau_R^2}{a}
    \Bigr[
    24\, b^{4} c^{3}\, \tau_m^{12} + 8\, b^{3} c^{2}\, \tau_m^{11} \tau_R (186 + 60\, \tau_R^{2} \Omega^{2} + \tau_m^{2} (227 + 95\, \tau_R^{2} \Omega^{2})\, \omega_{0}^{2} ) + 2\, b^{2} c\, \tau_m^{10} \tau_R^{2}\,
    \\
    &\times
    (3(6708 + 5405\, \tau_R^{2} \Omega^{2} + 1210\, \tau_R^{4} \Omega^{4} + 65\, \tau_R^{6} \Omega^{6}) + 4\, \tau_m^{2}\bigl(6390 + 5669\, \tau_R^{2} \Omega^{2} + 1198\, \tau_R^{4} \Omega^{4} + 11\, \tau_R^{6} \Omega^{6} )\, \omega_{0}^{2}) 
    \\
    & + \tau_m \tau_R^{4}
    (d
    )
    + 2c [108\,\tau_R^{12} + 3\,\tau_m^{4}\tau_R^{8}\,(56007 + 17351\,\tau_R^{2}\Omega^{2} + 632\,\tau_R^{4}\Omega^{4})
    + 6\,\tau_m^{5}\tau_R^{7}\,(50774 + 26631\,\tau_R^{2}\Omega^{2} + 2689\,\tau_R^{4}\Omega^{4})
    \\
    & + 3\,\tau_m^{8}\tau_R^{4}\,(65093 + 105663\,\tau_R^{2}\Omega^{2} + 52519\,\tau_R^{4}\Omega^{4} + 8821\,\tau_R^{6}\Omega^{6} + 328\,\tau_R^{8}\Omega^{8})
    \\
    & + 16\,\tau_m^{14}\,
(1 + \tau_R^{2}(\Omega - 3\omega_{0})^{2})
(4 + \tau_R^{2}(\Omega - \omega_{0})^{2})
(\omega_{0} + \tau_R^{2}\Omega^{2}\omega_{0} + \tau_R^{2}\omega_{0}^{3})^{2} (4 + \tau_R^{2}(\Omega + \omega_{0})^{2})
(1 + \tau_R^{2}(\Omega + 3\omega_{0})^{2})
]
\\
& + 2b\,\tau_m^{9}\tau_R^{3}[
6c\,(
13058
+ 15185\,\tau_R^{2}\Omega^{2}
+ 5068\,\tau_R^{4}\Omega^{4}
+ 493\,\tau_R^{6}\Omega^{6})
\\
&
+ \tau_m^{2}\omega_{0}^{2}\,
(474748
+ 238032\,\tau_m^{2}\omega_{0}^{2}
+ \tau_R^{2}\Omega^{2}\,(693851
+ 335241\,\tau_R^{2}\Omega^{2}
+ 60481\,\tau_R^{4}\Omega^{4}
+ 3119\,\tau_R^{6}\Omega^{6}
+ 24\,\tau_R^{8}\Omega^{8})
\\
& + 4\,\tau_m^{2}\,(42433
- 2377\,\tau_R^{2}\Omega^{2}
- 5577\,\tau_R^{4}\Omega^{4}
- 827\,\tau_R^{6}\Omega^{6}
+ 8\,\tau_R^{8}\Omega^{8})
\,\omega_{0}^{2})]
    \Bigr]
    \Bigg].
\label{eq:r4_st}
\end{aligned}
\end{equation}
Where,
\begin{align}
a \;=&\; \nonumber
\tau_m^{2}
(
    (4\tau_m + \tau_R)^{2}
    + 4 \tau_m^{2}\tau_R^{2}\Omega^{2}
)
(
    3 + 4 \tau_m^{2}\omega_0^{2}
)
\\[6pt] \nonumber
&\times
(
    (1 + \tau_R^{2}\Omega^{2})
    ( (\tau_m + \tau_R)^{2} + \tau_m^{2}\tau_R^{2}\Omega^{2} )
    + 2\tau_m \tau_R^{2}(\tau_m + \tau_R - \tau_m \tau_R^{2}\Omega^{2})\omega_0^{2}
    + \tau_m^{2}\tau_R^{4}\omega_0^{4}
)
\\[6pt] \nonumber
&\times
(
    (4 + \tau_R^{2}\Omega^{2})
    ( (2\tau_m + \tau_R)^{2} + \tau_m^{2}\tau_R^{2}\Omega^{2} )
    + 2\tau_m \tau_R^{2}\bigl(2\tau_R + \tau_m(4 - \tau_R^{2}\Omega^{2})\bigr)\omega_0^{2}
    + \tau_m^{2}\tau_R^{4}\omega_0^{4}
)
\\[6pt] \nonumber
&\times
(
    (1 + \tau_R^{2}\Omega^{2})
    ( (\tau_m + 3\tau_R)^{2} + \tau_m^{2}\tau_R^{2}\Omega^{2} )
    + 18\tau_m \tau_R^{2}(\tau_m + 3\tau_R - \tau_m\tau_R^{2}\Omega^{2})\omega_0^{2}
    + 81 \tau_m^{2}\tau_R^{4}\omega_0^{4}
)
\\[6pt] \nonumber
&\times
(
    12\tau_m \tau_R^{3}
    + 4\tau_R^{4}
    + \tau_m^{4}(
        1
        + \tau_R^{4}(\Omega^{2} - \omega_0^{2})^{2}
        + 2\tau_R^{2}(\Omega^{2} + \omega_0^{2})
    )
\\ 
&\qquad
    + 6\tau_m^{3}\bigl(\tau_R + \tau_R^{3}(\Omega^{2} + \omega_0^{2})\bigr)
    + \tau_m^{2}\tau_R^{2}(
        13 + \tau_R^{2}(5\Omega^{2} + 4\omega_0^{2})
    ))
\\
b =& \; 1 + \tau_R^2 \Omega^2
\\
c=& \; 4 + \tau_R^2 \Omega^2
\\
d =& \;   \nonumber
    6 \tau_R (4 + \tau_R^2 \Omega^2) (636 \tau_R^6 + \tau_m \tau_R^5 (4813 + 265 \tau_R^2 \Omega^2) + \tau_m^2 (20642 \tau_R^4 + 3230 \tau_R^6 \Omega^2) 
    \\ \nonumber
    &+
    2 \tau_m^6 (54633 + 64175 \tau_R^2 \Omega^2 + 20891 \tau_R^4 \Omega^4 + 1845 \tau_R^6 \Omega^6))
    +2 (1561268 \tau_m^9 + 3656211 \tau_m^8 \tau_R + 
6061622 \tau_m^7 \tau_R^2 
\\ \nonumber
&+ 7079545 \tau_m^6 \tau_R^3 + 
5783340 \tau_m^5 \tau_R^4 + 3255233 \tau_m^4 \tau_R^5 + 
1225558 \tau_m^3 \tau_R^6 + 291867 \tau_m^2 \tau_R^7 + 
39348 \tau_m \tau_R^8 + 2268 \tau_R^9
\\ \nonumber
&+ \tau_R^2 (2983088 \tau_m^9 + 5364109 \tau_m^8 \tau_R + 
6689016 \tau_m^7 \tau_R^2 + 5692480 \tau_m^6 \tau_R^3 + 
3252368 \tau_m^5 \tau_R^4 + 1219501 \tau_m^4 \tau_R^5 
\\ \nonumber
&+ 
289988 \tau_m^3 \tau_R^6 + 41871 \tau_m^2 \tau_R^7 + 
3636 \tau_m \tau_R^8 + 216 \tau_R^9) \Omega^2
+ \tau_m^2 \tau_R^4 (1937326 \tau_m^7 + 2522717 \tau_m^6 \tau_R 
\\ \nonumber
&+ 
2203240 \tau_m^5 \tau_R^2 + 1254095 \tau_m^4 \tau_R^3 + 
454498 \tau_m^3 \tau_R^4 + 103088 \tau_m^2 \tau_R^5 + 
14722 \tau_m \tau_R^6 + 1374 \tau_R^7) \Omega^4
\\ \nonumber
&+ \tau_m^4 \tau_R^6 (502910 \tau_m^5 + 428221 \tau_m^4 \tau_R + 
232000 \tau_m^3 \tau_R^2 + 77978 \tau_m^2 \tau_R^3 + 
16226 \tau_m \tau_R^4 + 2418 \tau_R^5) \Omega^6 )
\\ \nonumber
& + 2 \tau_m^6 \tau_R^8 (22375 \tau_m^3 + 10356 \tau_m^2 \tau_R + 2717 \tau_m \tau_R^2 + 801 \tau_R^3) \Omega^8 + 2 \tau_m^8 \tau_R^{10} (161 \tau_m + 183 \tau_R) \Omega^{10} ) \omega_0^2
\\ \nonumber
&+ \tau_m (2417520 \tau_m^{10} + 7385158 \tau_m^9 \tau_R + 15307381 \tau_m^8 \tau_R^2 + 22727848 \tau_m^7 \tau_R^3 + 24539711 \tau_m^6 \tau_R^4 + 19096738 \tau_m^5 \tau_R^5 
\\ \nonumber
&+ 10463263 \tau_m^4 \tau_R^6 + 3889860 \tau_m^3 \tau_R^7 + 923385 \tau_m^2 \tau_R^8 + 124812 \tau_m \tau_R^9 + 7236 \tau_R^{10} + 2 \tau_R^2 (1837500 \tau_m^{10} 
\\ \nonumber
&+ 4954700 \tau_m^9 \tau_R + 8653936 \tau_m^8 \tau_R^2 + 10118116 \tau_m^7 \tau_R^3 + 8016876 \tau_m^6 \tau_R^4 + 4256045 \tau_m^5 \tau_R^5 + 1452803 \tau_m^4 \tau_R^6 
\\ \nonumber
&+ 293381 \tau_m^3 \tau_R^7 + 29103 \tau_m^2 \tau_R^8 + 810 \tau_m \tau_R^9 + 54 \tau_R^{10}) \Omega^2 + \tau_m^2 \tau_R^4 (1413600 \tau_m^8 + 3466100 \tau_m^7 \tau_R + 5254694 \tau_m^6 \tau_R^2 
\\ \nonumber
&+ 4850010 \tau_m^5 \tau_R^3 + 2650780 \tau_m^4 \tau_R^4 + 814888 \tau_m^3 \tau_R^5 + 119895 \tau_m^2 \tau_R^6 + 5326 \tau_m \tau_R^7 + 579 \tau_R^8) \Omega^4 + 2 \tau_m^4 \tau_R^6 (46296 \tau_m^6 
\\ \nonumber
&+ 187812 \tau_m^5 \tau_R + 248162 \tau_m^4 \tau_R^2 + 146404 \tau_m^3 \tau_R^3 + 33711 \tau_m^2 \tau_R^4 + 1100 \tau_m \tau_R^5 + 315 \tau_R^6) \Omega^6 + \tau_m^6 \tau_R^8 (-22800 \tau_m^4 
\\ \nonumber
&+ 4070 \tau_m^3 \tau_R + 7221 \tau_m^2 \tau_R^2 - 1074 \tau_m \tau_R^3 + 171 \tau_R^4) \Omega^8 + 4 \tau_m^8 \tau_R^{10} (-186 \tau_m^2 + 16 \tau_m \tau_R + 3 \tau_R^2) \Omega^{10}) \omega_0^4
\\ \nonumber
&+ 2 \tau_m^2 \tau_R (1136616 \tau_m^{10} + 4188134 \tau_m^9 \tau_R + 8821012 \tau_m^8 \tau_R^2 + 12141082 \tau_m^7 \tau_R^3 + 11768317 \tau_m^6 \tau_R^4 + 8277592 \tau_m^5 \tau_R^5 
\\ \nonumber
&+ 4178170 \tau_m^4 \tau_R^6 + 1450862 \tau_m^3 \tau_R^7 + 323697 \tau_m^2 \tau_R^8 + 41274 \tau_m \tau_R^9 + 2268 \tau_R^{10} + \tau_m \tau_R^2 (1796384 \tau_m^9 + 5387784 \tau_m^8 \tau_R 
\\ \nonumber
&+ 9195742 \tau_m^7 \tau_R^2 + 9714228 \tau_m^6 \tau_R^3 + 6679451 \tau_m^5 \tau_R^4 + 3055202 \tau_m^4 \tau_R^5 + 911123 \tau_m^3 \tau_R^6 + 162910 \tau_m^2 \tau_R^7 
\\ \nonumber
&+ 13422 \tau_m \tau_R^8 + 18 \tau_R^9) \Omega^2 + 2 \tau_m^3 \tau_R^4 (426332 \tau_m^7 + 873194 \tau_m^6 \tau_R + 1014454 \tau_m^5 \tau_R^2 + 707163 \tau_m^4 \tau_R^3 + 307058 \tau_m^3 \tau_R^4 
\\ \nonumber
&+ 80144 \tau_m^2 \tau_R^5 + 9065 \tau_m \tau_R^6 + 7 \tau_R^7) \Omega^4 + 2 \tau_m^5 \tau_R^6 (69648 \tau_m^5 + 82524 \tau_m^4 \tau_R + 64423 \tau_m^3 \tau_R^2 + 25685 \tau_m^2 \tau_R^3 
\\
&+ 2880 \tau_m \tau_R^4 - 117 \tau_R^5) \Omega^6 + 2 \tau_m^7 \tau_R^8 (4000 \tau_m^3 + 175 \tau_m^2 \tau_R - 606 \tau_m \tau_R^2 - 3 \tau_R^3) \Omega^8 + 8 \tau_m^9 \tau_R^{10} (-76 \tau_m + \tau_R) \Omega^{10}) \omega_0^6
\end{align}

Finally, $\kappa_r$ can be calculated by substituting Eqs.\eqref{eq:msd_st} and \eqref{eq:r4_st} in Eq.~\eqref{eq:kurtosis_def}. 

\end{widetext}

\end{document}